\documentclass[11pt]{article} 
\usepackage{amsfonts, amssymb, amsmath,color, lineno, enumitem}
\usepackage[colorinlistoftodos]{todonotes}

\def\be{\begin{eqnarray}}
\def\ee{\end{eqnarray}}
\def\bee{\begin{eqnarray*}}
\def\eee{\end{eqnarray*}}
 \def\pmx{\begin{pmatrix}}
 \def\emx{\end{pmatrix}}
 \def\bsq{\begin{subequations}}
\def\esq{\end{subequations}}

\newtheorem{thm}{Theorem}

\newtheorem{lemma}[thm]{Lemma}

        \def\tr{\hbox{\rm Tr} \, }
          
          \def\trp{\hbox{\rm Tr} }
     
     \def\half{{\textstyle \frac{1}{2}}}
     \def\nn{\nonumber}

\def\hil{{\cal H}}

\def\eps{\epsilon}

\def\ds{\displaystyle}
 \def\ts{\textstyle}
\def\bra{\langle}
\def\ket{\rangle}
\def\kb{ \ket \bra }

\def\rt2{ \frac{1}{\sqrt{2}} }

\def\raw{\rightarrow}

           \def\wtd{\widetilde}
\newcommand{\proj}[1]{ | #1 \kb  #1|}
\newcommand{\ovb}[1]{\overline{ #1 }}
   \def\qed{ \hfill QED}

\def\ot{\otimes}

\title{Yet Another Proof of the Joint Convexity of Relative Entropy}  % \\ and Related Inequalities}

  \author{Mary Beth Ruskai \\
  Department of Mathematics \\ University of Vermont  \\ Burlington, VT 05405 USA \\ {\small mbruskai@gmail.com} }

\date{ { {\it \large Dedicated to the memory of Derek W. Robinson }} \\  ~~ \\ \today}

\begin{document}

\maketitle

\begin{abstract}
The joint convexity of the map $(X,A) \mapsto X^* A^{-1} X$, an integral representation of  operator convex functions,
and an observation of Ando are used to obtain a simple proof of both the joint convexity of relative entropy
and a trace convexity result of Lieb.  The latter was the key ingredient in the original proof of the strong subadditivity
of quantum entropy.
\end{abstract}

    %  \pagebreak
 
\section{Introduction}

In their influential book {\it Quantum Computation and Quantum Information}, Nielsen and Chuang \cite[Appendix 6]{NC}
assert  that ``no transparent proof of strong subadditivity'' of quantum entropy is known.
Since then there have been a number of simpler proofs, e.g., \cite{E,Rsimp} as well as more transparent
expositions \cite{C,Review,RsimpLb} of earlier arguments, including Lieb's key result \cite{Lb} on the joint concavity of the map
$(A,B) \mapsto \tr K^* A^p K B^{1-p} $ for $p \in (0,1) $ and $A, B \geq 0$ positive semi-definite matrices.
Indeed,   Simon's recent book \cite{S} on Loewner's Theorem includes three different proofs of Lieb's result! 

When Lieb and the author were working on strong subadditivity  (SSA) in 1971-72, they proved \cite{LbR3} that the 
map  $(X, A) \mapsto X^* A^{-1} X $ is jointly convex in $(X,A) $ with $A$ positive.\footnote{While  
referereeing a paper in  2010, the author was surprised to read that
``the joint convexity of $(X, A) \mapsto X^* A^{-1} X $
was proved by Kiefer in 1959 and rediscovered by Lieb and Ruskai''. We learned this just in time to add a
 reference to Kiefer \cite[Lemma~3.2]{K} to the final version of \cite{JR}.}
 They were able to use this result to prove a number of special cases of SSA,
but it seemed insufficient  for the general result.   Later, the author realized that the {\em strategy} 
used in \cite{LbR3} could also be used to prove   related results, e.g., 
 \cite[Appendix]{JR} and  \cite[Section~2.4]{Rsimp}, which lead to simple proofs of SSA.
%In particular, for $A, B > 0$  let $L_A$ and $R_B$ denote the left and right multiplication 
%operators  $L_A(X)\equiv  AX$ and $ R_B(X) \equiv XB $.  Then the map
%$(A,B, X) \mapsto   \linebreak  \tr X^*( L_A + t R_B )^{-1} X$ is jointly convex in $A,B$ when $t > 0 $. 
% However,
%this is an example of a convex trace function, not a result about operator convexity.

A few years after Lieb's paper \cite{Lb}, Ando \cite{A78,A79} gave two different proofs of Lieb's  key result.  After reading a draft of \cite{S}, the author 
realized that Ando's tensor product approach could be combined with the joint convexity \cite{K,LbR3}  of $(X, A) \mapsto X^* A^{-1} X $ to  prove 
the joint convexity of relative entropy  as well as Lieb's  result \cite{Lb}.
Historically, Lieb's concavity result was a key step in the first proof of SSA.  However, by proving
 the joint convexity of relative entropy directly \cite{JR,Rsimp}  one can  
 prove SSA  without Lieb's result.
In this note, we present a concise proof of both Lieb's result and the joint convexity of relative entropy.
We conclude by using the latter to give a simple proof of the monotonicity of relative entropy
under
partial traces and, hence, SSA. 

\section{Preliminaries} 

A mixed state for a quantum system associated with a finite dimensional Hilbert space $\hil$  is given by
 a density matrix, i.e., a positive semi-definite matrix $\rho $ with $\tr \rho = 1$.
The quantum entropy of a state $\rho$ was defined by von Neumann as $S(\rho) \equiv - \tr \rho \log \rho$.
Umegaki \cite{U} defined the relative entropy for a pair of states as 
$H(\rho,\gamma) \equiv \tr (\rho \log \rho - \rho \log \gamma)$.
Note that the relative entropy is well-defined if $\ker \gamma \subseteq \ker \rho$, and easily extended
to arbitrary matrices $A, B \geq 0 $.
Expositions of these concepts can be found in \cite{NC,OP,W}.

A quantum system with two (or more) subsystems is described by a tensor product of two (or more)
Hilbert spaces $\hil_{12} = \hil_1 \ot \hil_2 $.  If $\rho_{12} $ is a density matrix on $\hil_{12} $,
then the partial trace gives density matrices $\rho_1 = \trp_2 \, \rho_{12} $ and $\rho_2 = \trp_1 \, \rho_{12} $
on $\hil_1$ and $\hil_2$ respectively.  

The strong subadditivity (SSA) inequality for a state $\rho_{123}$  on $\hil_{123} = \hil_1 \ot \hil_2 \ot \hil_3 $ is
\be   \label{ssa} 
    S(\rho_{123} ) + S(\rho_2)   \leq S(\rho_{12} )  + S(\rho_{23} )  ~.
\ee
SSA was conjectured by Lanford and Robinson \cite{LR} in 1968 and proved by Lieb and Ruskai \cite{LbR1,LbR2} in 1973.	
Equivalent formulations can be found in \cite{LbR2,OP,Review,W}.

Our argument relies heavily on well-known results about operator convex functions.
A function  $f:(a,b) \mapsto {\bf R} $  is operator monotone if $A  \leq B \Rightarrow f(A) \leq  f(B) $ for self-adjoint $A,B$.
It is operator convex if $f\big(t A + (1-t) B \big) \leq  t f(A) + (1-t) f(B) $ holds  as an operator inequality for $t \in [0,1] $.
\linebreak  A function  $f(x) $ is operator monotone
 if it can be analytically continued to the upper half plane and  
maps the  upper half plane into the upper half plane  \cite[Theorem 2.7]{S}. It is operator convex  if 
a suitable   difference  quotient is operator monotone \cite[Theorem 9.1]{S}.  

\begin{thm} \label{ghdef}
Define $g_p(x) $ and $h_q(x) $  on $(0,\infty)$ by
\be
g_p(x) \equiv   \begin{cases} \tfrac{1}{1-p} ( x - x^p)   &  p \in (0,1) \cup (1,2] \\  x \log x & p = 1 \end{cases}  \qquad
h_q(x) \equiv \begin{cases}  \tfrac{1}{q}  (1 - x^q)& q \in [-1,0) \cup (0,1) \\  - \log x & q = 0 \end{cases}.
\ee
\end{thm} 
Then $g_p$ and $h_q $ are operator convex for $p \in (0,2] $ and $q \in [-1,1) $ respectively.

{\bf Proof:} The function $ f(x) = x \log x $ is operator convex
because $\frac{f(x) - f(0) }{x - 0 } = \log x $ is operator monotone.   
Since $f(x) = x $ is linear, the operator convexity of $g_p(x) $ 
depends on the behavior of $x^p$.    Observe that  $x^t$ is operator
monotone for $t \in (0,1) $ and  $-x^t $ is operator monotone for  $t \in [-1,0] $
so that $ x^p$ is operator convex for $p \in (1,2]$ and $-x^p$ is operator convex for $p \in (0,1) $.
Combined with the sign change   at $p = 1$, this implies that $g_p(x)$ is operator convex.
 
Next, observe that $h_q(x) = x g_{1-q}(x^{-1} )$.    When $g$  is operator convex  and $g(1) = 0 $,  then $\frac{ g(x) }{ x-1} $ is operator monotone.  Since $ x^{-1}  $ maps the upper half plane to the lower half plane,  
$\tfrac{ x g(x^{-1} ) }{ x-1} =  \tfrac{ - \, g(x^{-1})}{   x^{-1} -1 } $ is  operator monotone  and $x g(x^{-1} ) $
is operator convex.  Thus $h_q(x) $ is operator convex   for $q \in  [-1,0) \cup (0,1)  $.
\qed
 
\section{Joint convexity framework}

\begin{thm} \label{thm:kief} {\rm (Kiefer \cite[Lemma~3.2]{K} )}
Let $A$ be positive semi-definite with $\ker A \subseteq \ker X X^* $.  Then the map
$(X, A) \mapsto X^* A^{-1} X$ is jointly convex in $X,A$.
\end{thm} 

{\bf Proof:} Kiefer considered $A _j > 0 $ positive definite.   We give an argument 
based on  \cite{LbR3}.  
%  First, observe that since $(t X)^* ( tA)^{-1} ( t X ) = t (X^* A^{-1} X $ it suffices to show subadditivity.
  Let $M_j =t_j^{1/2} \big( A_j^{-1/2} X_j -  A_j^{1/2} \Lambda  \big) $ and
$\Lambda = \big( \sum_j t_j A_j \big)^{-1}  \big( \sum_j t_j X_j \big)$  with $t_j > 0$ and $  \sum_j t_j = 1$.  Then
\be
  0  ~  \leq  ~  \ts{ \sum_j } M_j^* M_j & = & \ts{  \sum_j  t_j \, X_j^* A_j^{-1} X_j  - \big( \sum_j  t_j X_j^* \big) \Lambda
          - \Lambda^* \big( \sum_j t_j X_j  \big)  + \Lambda^*\big( \sum_j t_j A_j  \big) \, \Lambda  }  \nn \\
          & = & \ts{  \sum_j t_j \, X_j^* A_j ^{-1} X_j  -  \big( \sum_j t_j X_j^* \big)  \big( \sum_j t_j A_j \big)^{-1} \big( \sum_j t_j X_j }\big)  
\ee
which proves joint convexity for $A_j > 0 $.  If some $A_j \geq 0 $ is singular and $\ker A_j \subseteq \ker X_j X_j^* $, then
   $\ds{\lim_{\eps \raw 0}} X_j^* \big(A_j + \eps I \big) X_j $ exists.  Thus, one can repeat the argument above
   with $A_j$ replaced by $A_j + \eps I $ and take the limit $\eps \raw 0 $. \qed

Henceforth, for simplicity, we consider only positive definite matrices $A, B > 0 $.
The reader can readily discern the conditions under which some results extend to 
positive semi-definite matrices.   

\begin{lemma}{\rm (Ando \cite{A78,A79})} \label{thm:ando}
 Let $A, B$ be $m \times m$ and $n \times n$ matrices respectively and let
$K$ be an $m \times n$ matrix considered as a vector in ${\bf C}_m \ot {\bf C}_n $.  Then
\be   \label{ando}
   \tr K^* A K B = \bra K ,(A \ot B^T) K \ket ~.
\ee
\end{lemma}

\begin{thm} \label{thm:main}
Let $g : (0, \infty) \mapsto {\bf R} $ be an operator  convex function with $g(1) = 0$ .  Then for $A, B > 0 $
 the map
$(A, B) \mapsto   (I \ot B)  g(A \ot  B^{-1} ) $ is jointly  convex.
\end{thm} 
{\bf Proof:} It follows from \cite[Example 12.8]{S}  that the function $g(x) $ can be written in the form 
\be \label{intrep}
  g(x) = b\,  (x-1) + c\,  (x-1)^2 + \int_0^\infty \frac{(x-1)^2}{x+s} d\mu(s) 
\ee
with $c  \geq 0$ and  $\mu(s) $ a positive measure on $(0,\infty)$ with $\int_0^\infty \tfrac{1}{1+s} d\mu(s)  < \infty$ .  Then
\be  
(I \ot B)  g(A \ot B^{-1} ) & = &   b \,  (A \ot I - I \ot B )    + c \,  (\, A \ot I - I \ot B )\frac{1}{  I \ot B } (A \ot I - I \ot B )\quad \nn  \\
  & ~ &      + \int_0^\infty  (A \ot I - I \ot B) \frac{1}{A \ot I + s I \ot B } (A \ot I - I \ot B  ) \,  d\mu(s) 
\ee
The first term is linear in $A,B$; and the remaining terms are jointly convex in $A,B$ by Theorem~\ref{thm:kief}.
Since $c \geq 0 $ and $\mu$ is a positive measure, the result follows.   \qed
\medskip

\noindent{\bf Remark:}   $B \mapsto B^T$ is linear, $B \mapsto \ovb{B}$ is affine and   $B > 0 $ implies $ B^T = \ovb{B} $.
Therefore, $(A, B) \mapsto   (I \ot \ovb{B}) g(A \ot  \ovb{B}^{-1} )  $ is also jointly convex in  $ (A,B) $.
Thus, in the applications which follow, one can  replace $B^T$ by  $B$ on the right in \eqref{ando}.

\section{Joint convexity examples}  \label{sect:examp}

We now apply these results to the functions defined in  Theorem~\ref{ghdef} to obtain the joint convexity results in \cite{LbR2}
and \cite{Lb} important for proving SSA. 
\begin{enumerate}[label=\alph{*})]

 \item   For $ p \in (0,1) \cup (1,2]$ the map
 $ (A,B) \mapsto (I \ot \ovb{B})  \, g_p(A \ot  \ovb{B}^{-1} )= \tfrac{1}{1-p} \big( A \ot I - A^p \ot \ovb{B}^{1-p} \big)$ 
 is jointly convex in $A,B > 0 $
      which implies that  joint convexity also holds for  
      $$  (A, B) \mapsto \tfrac{1}{1-p}  \tr \big( K^* A K - K^* A^p K B^{1-p} \big) ~. $$ 
      
      \item   The map $(A,B) \mapsto (I \ot \ovb{B})  (A \ot  \ovb{B}^{-1} ) \log (A \ot  \ovb{B}^{-1} )  = 
      (A \ot I)\big[  \log A \ot I - I \ot \log \ovb{B} \big] $  is jointly convex in $A,B > 0$
      which implies that   joint convexity also holds for  
       $$(A, B) \mapsto \tr \big(   K^*( A \log A) \, K - K^* A \, K \log B \big) ~. $$
      
      \item For $ q \in [-1,0) \cup (0,1)$ the map  $(A,B) \mapsto  ( I \ot \ovb{B})\,  h_q(A \ot  \ovb{B}^{-1} ) =  \tfrac{1}{q } \big( I  \ot \ovb{B}  - A^q  \ot \ovb{B}^{1-q} \big)$   is jointly convex in $A,B > 0$
      which implies that   joint convexity also holds for  
    $$ (A, B) \mapsto \tfrac{1}{q } \ \tr \big( K^*K B - K^* A^q K B^{1-q}\big)  ~. $$
      
      \item  The map $(A,B) \mapsto   -  (I \ot \ovb{B})  \log (A \ot  \ovb{B}^{-1} ) = (I \ot \ovb{B} ) \big(I \ot  \log \ovb{B}  -  \log A \ot I \big) $        is jointly convex in $A,B > 0 $     which implies that   joint convexity also holds for  
   $$(A, B) \mapsto \tr \big(   K^* K (B \log B ) -      K^*( \log A )K   B  \big)  ~.$$  

\end{enumerate}

      \section{Discussion} 
      
      Note that (a) and (c) imply that $\tfrac{1}{p(1-p) } \tr K^* A^p K B^{1-p} $ is jointly concave in $A,B > 0$ for all $p \in [-1,2]$ (extended by continuity at $p = 0,1$).  Thus,   $ p \in (0,1) $ gives
    Theorem 1  in \cite{Lb} when $r = 1-p$.   Ando \cite{A} also showed that $\tr K^* A^p K B^{1-p} $  is jointly convex   for $p \in [1,2]$.  
    Hasegawa \cite{H} seems to have been the first to realize that Lieb and Ando's results  can be extended to all $p \in [-1,2]$ in the form 
    stated here, as also observed in \cite{JR}.    
      
    The choice $K = I $ in (b) gives the joint convexity of the relative entropy $H(A,B) $,
     while $K = I $ in (d) gives the joint convexity of $H(B,A) $.
     More generally,   $ (A \ot I) \,  g_p( A^{-1} \ot B ) = (I \ot B)  \, h_{1-p} (A \ot  \nobreak B^{-1} ) $ so that (c) and (d) are somewhat redundant.
     However, using both $g_p(x) $ and  $ h_q(x) $ makes clear how the results vary for subintervals $ [0,1), (-1,0), (1, 2] $.
     
     The linear term $\tr K^* A K $ in (a) arises because $g_p(x) $ was defined so that $\lim_{ p \raw 1}\,  g_p(x) = x \log x$.
     Moreover,  the  Wigner-Yanase-Dyson   (WYD) entropy \cite{WY} for  a density matrix $\gamma$ and $K = K^* $,  was
  defined as   $ \half \tr  [K, \gamma^p][K, \gamma^{1-p} ] = \tr K \gamma^p K \gamma^{1-p}  - \tr K \gamma K $,  which
    contains a linear term.  Wigner and Yanase proved concavity  in  
     $ \gamma$  at $ p = \half$,  and Dyson suggested the extension to $p \in (0,1) $.  After dropping the linear term,
     Lieb \cite{Lb2}  proved a generalization of the WYD conjecture as described above.  The expressions  in (a) and (c) 
     can be regarded as a WYD version of relative entropy.
      
       In most applications, one chooses $K = I $.     However, 
       Kim \cite{Kim} showed \footnote{Kim's paper \cite{Kim} was posted on arxiv.org
     very close the the 40th anniversary  of the actual completion of the proof of SSA  in October, 1972.} 
    that by
   using a  judicious choice of  $K$ he could strengthen some 
      related inequalities to operator inequalities on one subspace of a tensor product.  In particular, using
      $K = I_1 \ot I_2 \ot \proj{\phi} $ with $ \phi \in \hil_3 $, he proved that
      \be
         \trp_{12}  \,  \rho_{123} \big[ \log \rho_{123} - \log \rho_{12} - \log \rho_{23} + \log \rho_2 \big] \geq 0
      \ee
     holds as an operator inequality on $\hil_3$.  For examples related to (a) and (c) see \cite{Rkim}.
     
     \section{Generalizations of Relative Entropy}  \label{sect:genent} 
     
     The examples above are special cases of a generalization of relative entropy \cite{HR,JR,LesR}  introduced 
     by Petz \cite{OP,PzQ}
     and sometimes called quasi-entropies or $f$-divergences.  These generalizations were
       defined using the relative modular operator $L_A R_B^{-1}  $   
       where $L_A(X) = AX $ and $R_B(X) = XB $ denote left and right multiplication.
   Let ${\cal G} = \{    g: (0,\infty)  \mapsto {\bf R}   :  g \hbox{ is operator convex and }  g(1) = 0  \} $.  Then for any $g \in {\cal G}$
   and   $A, B > 0$, define
     $H_g(K,A,B) = \tr K^*g(L_A R_B^{-1} ) R_B K$.    It follows from Lemma~\ref{thm:ando} that
   \be   \label{hsubg}
     H_g(K,A,B) =  \bra K ,  (I \ot \ovb{B} )  \, g [ (A \ot I) (I \ot \ovb{B}^{-1})]  K \ket  \, .
   \ee
Since $g(x) $ has an integral representation as in \eqref{intrep}, it 
  follows immediately from Theorem~\ref{thm:main} that   the  map
     $(A,B) \mapsto H_g(K,A,B) $ is jointly convex for all $g \in {\cal G}$.   
    
     Ando  proved \cite[Theorem~6]{A79} the joint
     concavity of   $(A,B) \mapsto   (I \ot \ovb{B} )  \, f [ (A \ot I) (I \ot \ovb{B}^{-1})] $  when $f$ 
      is a positive monotone operator function on $(0,\infty)$, and used this to prove Lieb's concavity result.  However,
      $f \notin {\cal G} $ because $f(1) \neq 0$.  He then \cite[Theorem~7]{A79}  extended this to arbitrary 
      monotone operator functions on $(0,\infty)$ and, in particular  \cite[Corollary~7.1]{A79}, 
         to $f(x) =  \log x $    which implies the joint
     convexity of relative entropy.
                 
      The argument used in the proof of Theorem~1 to show that $h_q(x) $ is operator convex 
      also  shows that   $\wtd{g}(x) = x g(x^{-1} ) $ is  operator convex  for all $g \in {\cal G}$.    Then
        $H_{\wtd{g}}(K,A,B) = H_g(K,B,A) $.  Moreover, the function
       $k(x) =[  g(x) + x g(x^{-1} )] (x-1)^{-2} $ is well-defined, operator convex, and operator monotone decreasing \cite{HR}.
     Petz \cite{PzR} showed that there is a one-to-one correspondence between $k(x) $ and Riemannian
     metrics which generalize the classical Fisher-Rao information and decrease under the action of  completely positive 
     trace-preserving   maps which describe quantum channels.   (See also \cite[Theorem II.14]{LR}.) 
          \section{Proof of strong subadditivity}            
      
      It is now well-known \cite{C, LbR2, Review, S, VCL, W}  that these convexity results can be used
       in a number of different ways to prove
      strong subadditivity of quantum entropy, as well as the fact that relative entropy decreases under partial traces and
     under the action of quantum channels, i.e., completely positive trace-preserving maps.  For completeness, we conclude
  by showing that joint convexity of $H(\rho,\gamma) $  implies monotonicity under partial traces and SSA.   \medskip
     
     To prove $H(\rho_1, \gamma_1 ) \leq  H(\rho_{12}, \gamma_{12} ) $ we use the the Weyl-Heisenberg  matrices
     $X$ and $Z$  in $M_d$ with elements  $x_{jk} = \delta_{j + 1, k} $ (with  $j + 1 \mod d$)  and 
     $z_{jk} = \omega^j \delta_{jk} $ where $\omega = e^{2 \pi i/ d} $.     Then  $X$ and $Z$
      are unitary  and for any matrix $P \in M_d$,
\be  \label{haar} 
   \sum_{m,n} X^m Z^n  P  Z^{-n} X^{-m} =  d \, ( \tr P ) I  ~.
   \ee
Now let $U_{mn} = I_1 \ot X^m Z^ n $   where $X, Z $  act on $\hil_2$.   Then 
$U_{mn} $ is a unitary matrix acting on $\hil_{12} $ and $U_{mn}^* = I_1 \ot Z^{-n} X^{-m} $.
It follows from \eqref{haar} that
$ \tfrac{1}{d_2^2} \sum_{m,n}   U_{mn} P_{12} U_{mn}^* = \tfrac{1}{d_2} P_1 \ot I_2 $.  
%(This is a pedestrian version of Uhlmann's observation  \cite{ } that the partial trace 
%can be represented as an integral of unitary conjugations over Haar measure.) 
Snce $H(\rho,\gamma) $ is invariant
under unitary conjugations,  its joint convexity implies  
      \be
   H(\rho_{12}, \gamma_{12} )  & = & 
        \sum_{m,n}      \tfrac{1}{d_2^2} H \big( U_{mn} \rho_{12} U_{mn}^*  \,   , \,   U_{mn} \gamma_{12} U_{mn}^* \big)  \nn \\
         & \geq &   H  \bigg(     \sum_{m,n}     \tfrac{1}{d_2^2} \, U_{mn} \rho_{12} U_{mn}^* ~ , ~    \sum_{m,n}  \tfrac{1}{d_2^2}  \, U_{mn} \gamma_{12} U_{mn}^* \bigg) \nn  \\ 
          & = & H  \big( \rho_1 \ot \tfrac{1}{d_2} I_2  \, , \,   \gamma_1 \ot \tfrac{1}{d_2} I_2 \big)  ~ = ~  H(\rho_1 \,  , \, \gamma_1) ~ .
 \ee
 Then onc can prove SSA by observing that
 \be
     S(\rho_2)  -   S(\rho_{12} )    & = &  H(\rho_{12} , \tfrac{1}{d_1} I_1 \ot \rho_2 ) + \log d_1    \nn  \\
                & \leq &  H(\rho_{123} , \tfrac{1}{d_1} I_1 \ot \rho_{23}  ) + \log d_1  \nn  \\
                & = &   S(\rho_{23} ) - S(\rho_{123} )
 \ee
which is equivalent to \eqref{ssa}.

\medskip
     
   \noindent{\bf Remark:} One can prove the joint convexity of   $H(A,B)$
   without reference to convex operator functions by using the elementary  integral representation  
   $\log x = \int_0^\infty \big( \frac{1}{ 1 + t} - \frac{1}{x + t} \big) dt $.  Then
   \be
          -  (I \ot \ovb{B} )   \log (A \ot B^{-1} )  & = &   (I \ot \ovb{B} )  \log  (I \ot \ovb{B} ) -  (I \ot \ovb{B} )   \log (A \ot I )   
           \nn   \\   
            & = &   \int_0^\infty \bigg(( I \ot \ovb{B}) \frac{1}{ (A \ot I ) + t (I \ot B) } ( I \ot \ovb{B}) - \frac{ I \ot \ovb{B} }{1 + t} \bigg) \, dt
   \ee
The first term is jointly convex in $A,B$ by Theorem~\ref{thm:kief}  and the second term is linear.     Thus,  Lemma~\ref{thm:ando} 
implies   the joint convexity of  relative entropy  as in Example~(d) of Section~\ref{sect:examp}.

           \bigskip
           
           \noindent{\bf Acknowledgment: } It is a pleasure to thank Professor Barry Simon for sending the author a
           draft of \cite{S} which reawakened the author's awareness of old results.   
      \bigskip
    
%    Data sharing not applicable to this article as no datasets were generated or analysed during the current study.
%    
%    The author has no conflicts of interest related to this manuscript.
 \pagebreak

\end{document}